\documentclass[%
 reprint,superscriptaddress,
 amsmath,amssymb,prl,aps,
]{revtex4-2}

\usepackage{graphicx}
\usepackage{dcolumn}
\usepackage{bm}
\usepackage{hyperref}
\usepackage{xcolor}

\def \kag{kagom{\'e}}

\begin{document}

\preprint{APS/123-QED}

\title{Damped Dirac magnon in a metallic kagom{\'e} antiferromagnet FeSn}

\author{Seung-Hwan Do}
\affiliation{Materials Science and Technology Division, Oak Ridge National Laboratory, Oak Ridge, Tennessee 37831, USA}

\author{Koji Kaneko}
\affiliation{Materials and Life Science Division, J-PARC Center, Tokai, Ibaraki, 319-1195, Japan}
\affiliation{Materials Sciences Research Center, Japan Atomic Energy Agency, Tokai, Ibaraki, 319-1195, Japan}

\author{Ryoichi Kajimoto}
\affiliation{Materials and Life Science Division, J-PARC Center, Tokai, Ibaraki, 319-1195, Japan}

\author{Kazuya Kamazawa}
\affiliation{Neutron Science and Technology Center, Comprehensive Research Organization for Science and Society, Tokai, Ibaraki, 319-1106, Japan}

\author{Matthew B. Stone}
\affiliation{Neutron Scattering Division, Oak Ridge National Laboratory, Oak Ridge, Tennessee 37831, USA}

\author{Shinichi Itoh}
\affiliation{Materials and Life Science Division, J-PARC Center, Tokai, Ibaraki, 319-1195, Japan}
\affiliation{Institute of Materials Structure Science, High Energy Accelerator Research Organization, Tsukuba, Ibaraki, 305-0081, Japan}

\author{Takatsugu Masuda}
\affiliation{Institute for Solid State Physics, The University of Tokyo, Chiba 277-8581, Japan}
\affiliation{Institute of Materials Structure Science, High Energy Accelerator Research Organization, Tsukuba, Ibaraki, 305-0081, Japan}

\author{German D. Samolyuk}
\affiliation{Materials Science and Technology Division, Oak Ridge National Laboratory, Oak Ridge, Tennessee 37831, USA}

\author{Elbio Dagotto}
\affiliation{Materials Science and Technology Division, Oak Ridge National Laboratory, Oak Ridge, Tennessee 37831, USA}
\affiliation{Department of Physics and Astronomy, University of Tennessee, Knoxville, Tennessee 37996, USA}

\author{William R. Meier}
\affiliation{Materials Science and Technology Division, Oak Ridge National Laboratory, Oak Ridge, Tennessee 37831, USA}

\author{Brian C. Sales}
\affiliation{Materials Science and Technology Division, Oak Ridge National Laboratory, Oak Ridge, Tennessee 37831, USA}

\author{Hu Miao}
\affiliation{Materials Science and Technology Division, Oak Ridge National Laboratory, Oak Ridge, Tennessee 37831, USA}

\author{Andrew D. Christianson}
\affiliation{Materials Science and Technology Division, Oak Ridge National Laboratory, Oak Ridge, Tennessee 37831, USA}


\begin{abstract}
The \kag{} lattice is a fertile platform to explore topological excitations with both Fermi-Dirac and Bose-Einstein statistics. While relativistic Dirac Fermions and flat-bands have been discovered in the electronic structure of \kag{} metals, the spin excitations have received less attention. Here we report inelastic neutron scattering studies of the prototypical \kag{} magnetic metal FeSn. The spectra display well-defined spin waves extending up to 120 meV. Above this energy, the spin waves become progressively broadened, reflecting interactions with the Stoner continuum. Using linear spin wave theory, we determine an effective spin Hamiltonian that reproduces the measured dispersion. This analysis indicates that the Dirac magnon at the K-point remarkably occurs on the brink of a region where well-defined spin waves become unobservable. Our results emphasize the influential role of itinerant carriers on the topological spin excitations of metallic \kag{} magnets.

\end{abstract}

This manuscript has been authored by UT-Battelle, LLC under Contract No. DE-AC05-00OR22725 with the U.S. Department of Energy.  The United States Government retains and the publisher, by accepting the article for publication, acknowledges that the United States Government retains a non-exclusive, paid-up, irrevocable, world-wide license to publish or reproduce the published form of this manuscript, or allow others to do so, for United States Government purposes.  The Department of Energy will provide public access to these results of federally sponsored research in accordance with the DOE Public Access Plan (http://energy.gov/downloads/doe-public-access-plan).

\maketitle

\begin{figure}[t!]
\includegraphics[width=8.5cm]{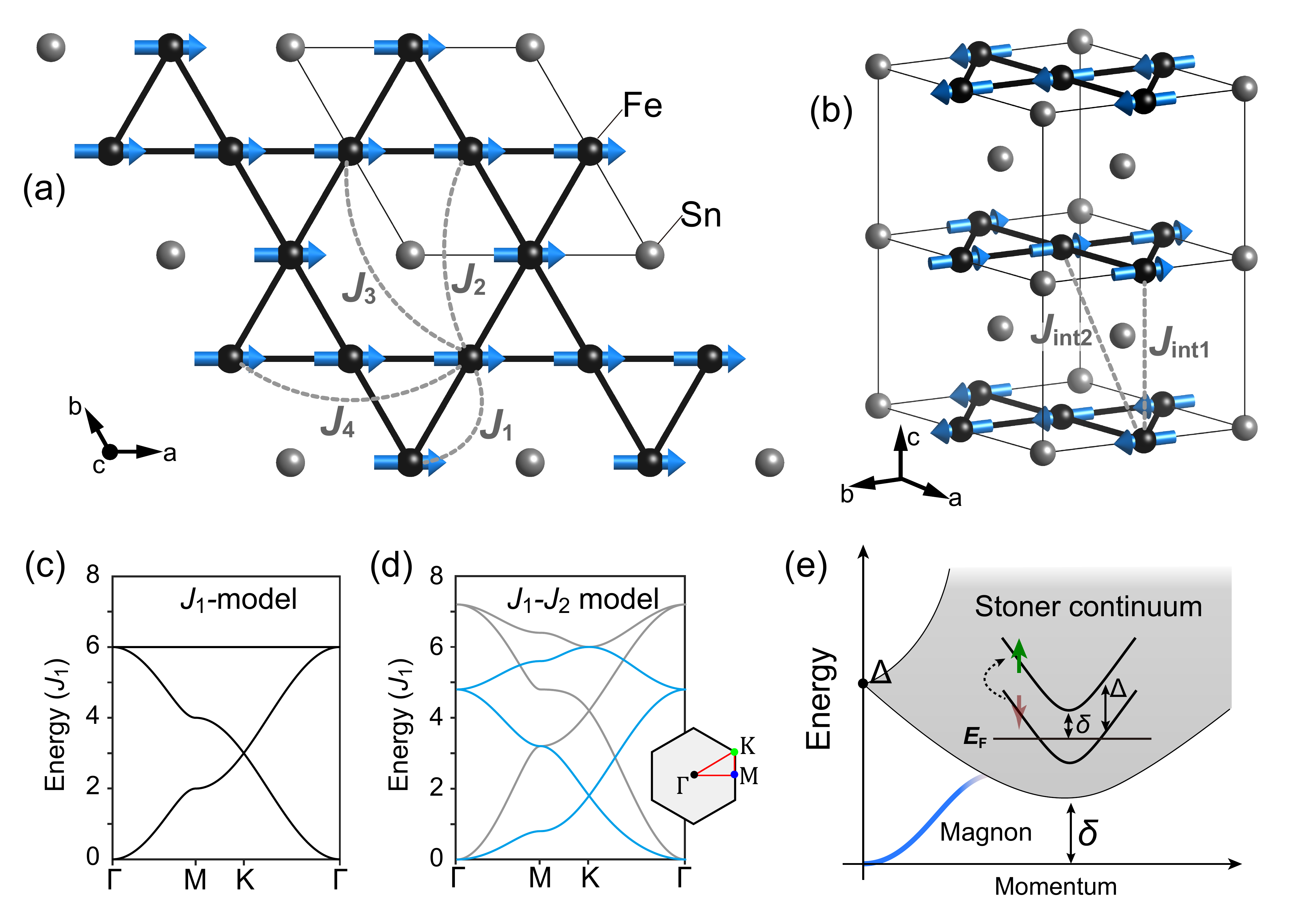} 
\caption{\label{fig:crystallography} 
(a) (b) Crystal and magnetic structure of FeSn. The exchange paths between Fe spins are indicated. The spin wave dispersions of a ferromagnetic \kag{} lattice with $J_1$=-1 meV and (c) $J_2$=0 and (d) $J_2$=0.2$J_1$ ($J_2$=-0.2$J_1$), are displayed with black and  gray (blue) curves, respectively. High symmetry points are indicated in the inset to (d). (e) Schematic of a Stoner excitation spectra (continuum) and magnon (sharp dispersion) as a function of momentum ($\textbf{Q}$) and energy ($E$). The spin wave mode decays into a particle-hole pair near the Fermi energy ($E_{F}$) when it enters the Stoner continuum. The continuum boundary shifts with gap $\Delta$ ($\delta$), reflecting the direct (indirect) electronic transitions, as shown in the inset. 
}
\end{figure}

The interplay between charge, spin, and geometric frustration is an important underlying theme to problems at the forefront of condensed matter physics~\cite{Tang2011,Sheng2011,Neupert2011,Sun2011, yin2018, ye2018, yin2019, kang_2020fesn, Yin2020}. Kagom{\'e} magnets, consisting of a corner shared transition-metal triangular-network (Fig.~\ref{fig:crystallography}(a)), are an ideal platform to explore correlated topological states, including the fractional quantum Hall effect \cite{Tang2011,Sheng2011,Neupert2011,Sun2011}, the intrinsic Chern state \cite{Thouless1982, Haldane1988, Xu2015, Yin2020} and magnetic Weyl semimetals \cite{Liu2019}. While the charge excitations of \kag{} magnets have been extensively investigated \cite{Kang2020CoSn, kang_2020fesn, Liu2020CoSn, ye2018, Ortiz2020, yin2018, yin2018, Yin2020, Liu2019}, their magnetic counterparts and the intertwined correlations between charge and spin degrees of freedom have not yet been investigated in detail. 

Similar to the electronic structure, a spin model with nearest-neighbor magnetic exchange, $J_{1}$, yields a Dirac magnon at the K-point and a flat magnon band, as shown in Fig.~\ref{fig:crystallography}(c)~\cite{owerre2017,mook2014magnon,chisnell2015}. Time-reversal symmetry breaking interactions, such as the Dzyaloshinskii-Moriya interaction in magnetic insulators, introduce a gap at the Dirac point and can induce a topological thermal Hall effect ~\cite{mook2014magnon,mook2014edge,chisnell2015,zhang2013,owerre2017}. In addition, magnon-magnon interactions may modify the dispersion to realize interaction-stabilized topological magnons~\cite{mook2021,mcClarty2019}. This simplified picture is, however, challenged in a metallic \kag{} magnet, where the presence of itinerant electrons will introduce long-range magnetic interactions through, $e.g.$ RKKY (Ruderman–Kittel–Kasuya–Yosida) interactions, that dramatically change the magnon dispersion as shown in Fig.~\ref{fig:crystallography}(d). Moreover, the high-energy spin wave excitations will interact with the particle-hole continuum of the Stoner excitation (Fig.~\ref{fig:crystallography}(e)), resulting in mode decay. 

To explore the effects of itinerant carriers on the magnons in the ferromagnetic \kag{} spin-lattice, we study the spin excitation spectra of the metallic \kag{} magnet FeSn using inelastic neutron scattering (INS). The measured spectra show relatively sharp spin waves of the ferromagnetic \kag{} spin-lattice below 120 meV. At higher energies, the spin waves exhibit decay due to interactions with the Stoner continua. Interestingly, we find that while the Dirac magnon remains, the upper branch of the Dirac band is heavily damped, uncovering a non-trivial interplay between magnon and continuum.

FeSn crystallizes in a hexagonal structure (P6$/mmm$) with the Fe atoms forming a two-dimensional \kag{} spin-lattice(Fig.~\ref{fig:crystallography}(b)). Below $T_\text{N}$=365 K, the Fe spins form ferromagnetic \kag{} layers which are stacked antiferromagnetically along the $c$-axis with an ordering wave vector of  $\text{\textbf{Q}}_\text{m}$=(0,0,1/2).  As we show in this letter, the dominant in-plane ferromagnetic interactions allow the behavior of the quasi two-dimensional ferromagnetic \kag{} spin-lattice to be probed. For the INS measurements, 4.43~g of FeSn single crystals were grown using the flux method~\cite{sales_2019} and co-aligned on aluminum plates with a [$H,0,L$] horizontal scattering plane. The INS data were obtained at $T$=100~K using HRC~\cite{itoh2011} (incident energies $E_\text{i}$=40 and 153 meV) and 4SEASONS~\cite{kajimoto2011} ($E_\text{i}$=27 meV, 46 meV, 96 meV, and $E_\text{i}$=300 meV) spectrometers at the Japan Proton Accelerator Research Complex (J-PARC). Additional data were collected with the SEQUOIA~\cite{granroth2010} spectrometer ($E_\text{i}$=500 meV) at the Spallation Neutron Source (SNS) at Oak Ridge National Laboratory (see \cite{supple} for additional details).

\begin{figure}[t]
\includegraphics[width=8.5cm]{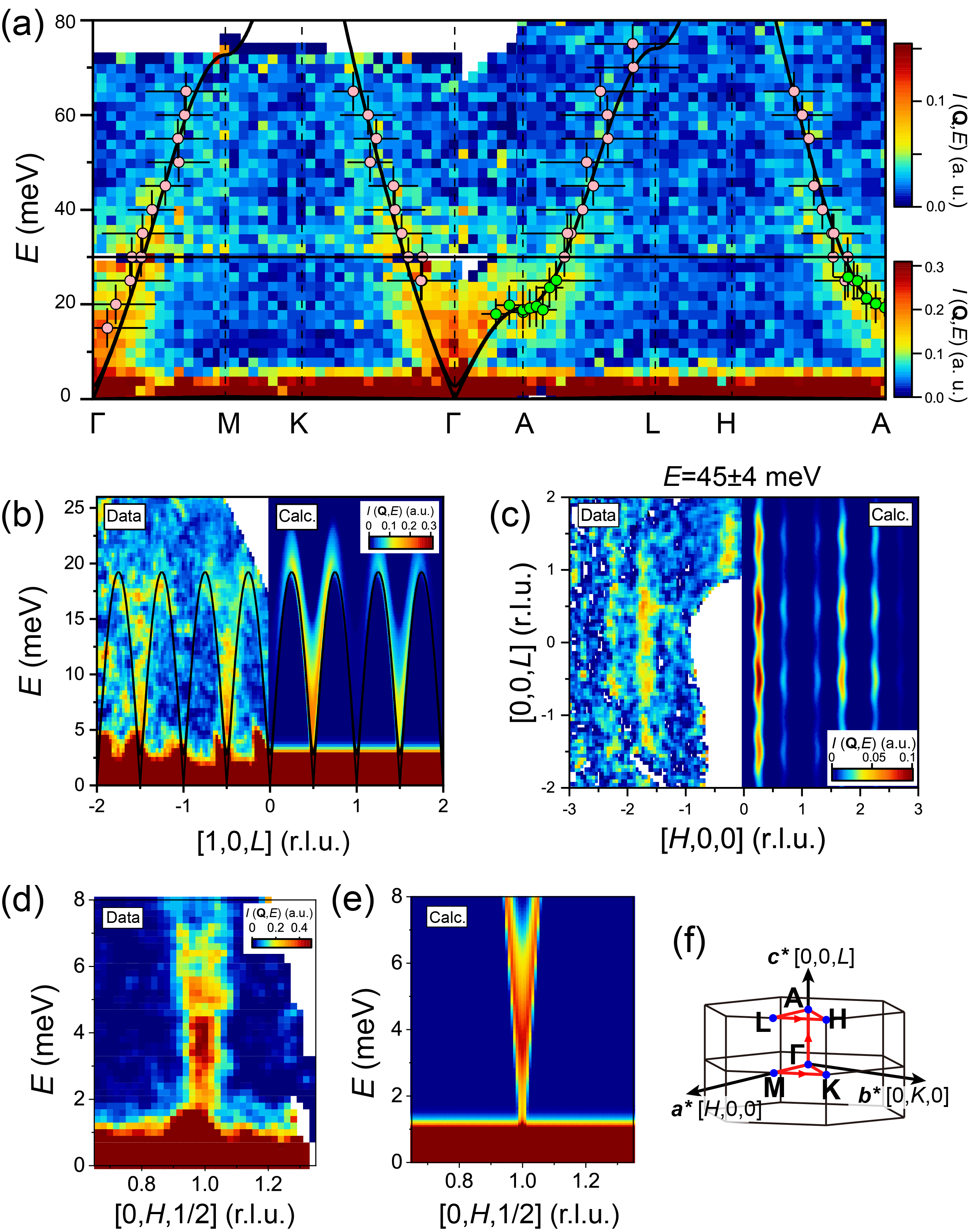} 
\caption{\label{fig:lowE_spec} (a) Contour map of the INS intensity along high symmetry directions (given in (f)).   The data ((a),(c)) were measured using HRC with $E_i$=153 meV. The spectrum above (below) the horizontal line at 30 meV was obtained from the BZ for $\Gamma$ at $\text{\textbf{Q}}$=(0,0,1/2) ((0,0,3/2)), integrating over  $\text{\textbf{Q}}$=0.22 \AA$^{-1}$ along the vertical direction. Horizontal (vertical) error bars of pink (green) circles indicate the fitted peaks full width at half maxima (FWHM), and vertical (horizontal) error bars indicate the range of energy (momentum) integration. (b) INS data (left) and spin wave calculations (right) as described in the text along the out-of-plane direction through the ZC, measured using the 4SEASONS spectrometer with $E_\text{i}$=46 meV. The solid line is the calculated magnon dispersion. (c) Constant energy slice of the magnon spectra in the [$H$,0,$L$] plane and the calculated spectra. (d) Low-energy spectrum of $I(\textbf{Q},E)$ near the ZC measured using $E_\text{i}$=27 meV at 4SEASONS, and (e) the corresponding calculation including an easy-plane anisotropy of $D_{z}$=0.2 meV.}
\end{figure}

\begin{figure}[t]
\includegraphics[width=8.5cm]{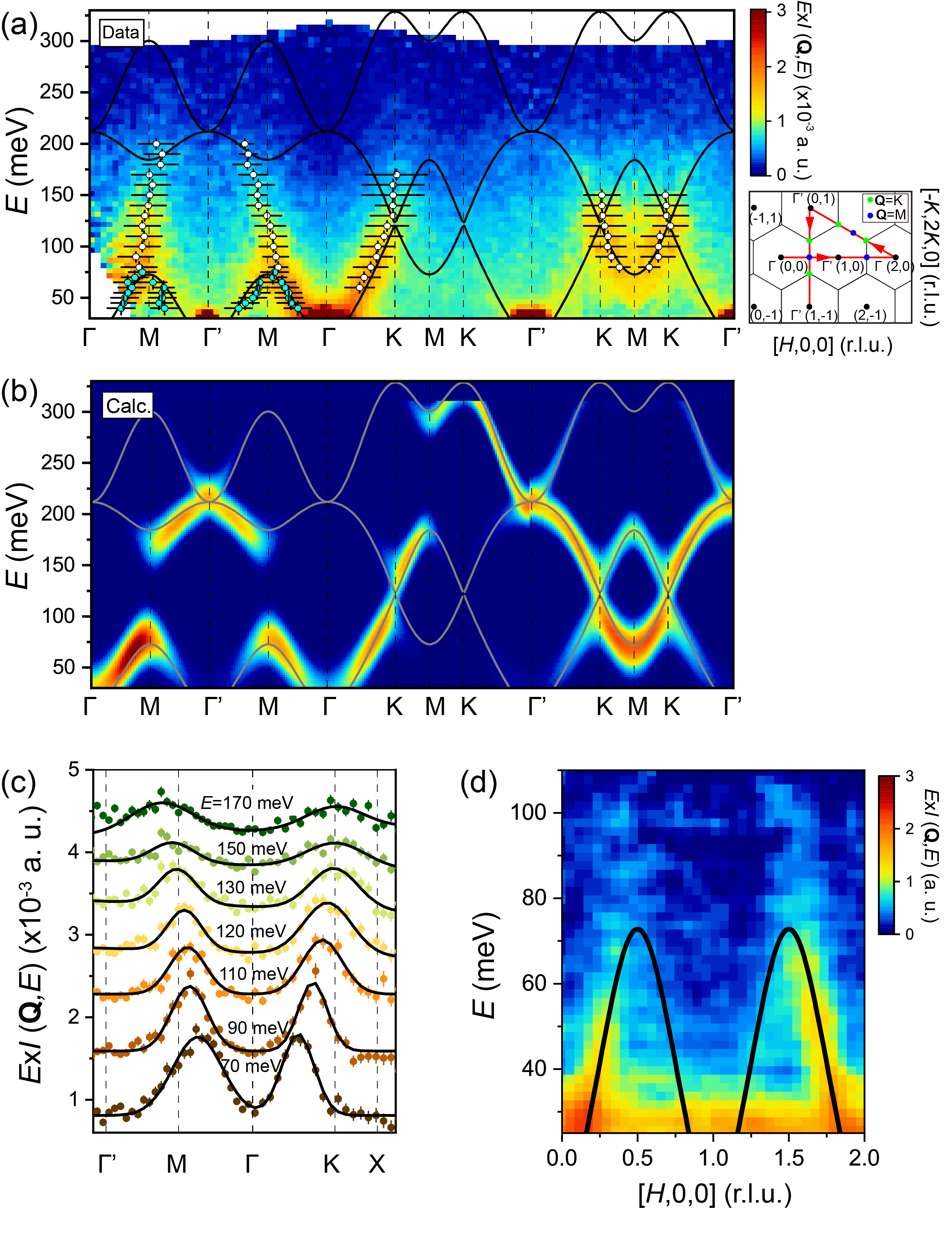}
\caption{\label{fig:highE_spec}
(a) High-energy INS spectra (plotted as $E\times I(\text{\textbf{Q}},E)$) and spin wave calculations ((b)) along high symmetry directions as indicated in the right panel of the $HK$-reciprocal space map.  Data were obtained by integrating over  $\text{\textbf{Q}}$=0.19 \AA$^{-1}$ and -4$\leq L \leq$4. The calculation was performed for an identical $\text{\textbf{Q}}$-integration range and convoluted with the instrumental resolution of SEQUOIA. The black solid lines display the magnon dispersion for $L$=0.5. 
Horizontal (vertical) error bars of white filled circles indicate the fitted peaks FWHM (range of energy integration). (c) Constant energy cut along the high symmetry directions, integrated over energy $\pm$5 meV. Solid lines are Gaussian fits described in the text with fitted values displayed in (a). (d) INS spectra obtained from HRC ($E_{i}$=153 meV), integrated over -3$\leq L \leq$3.  
}
\end{figure}

Figure~\ref{fig:lowE_spec} shows the spectra in the three-dimensional hexagonal Brillouin zone (BZ), measured by INS. The acoustic magnons emanate from $\textbf{Q}_{m}$=$\Gamma$(0,0,1/2), and disperse throughout the entire BZ. Strongly dispersive magnons in the $HK$-plane extend well above 80 meV, whereas the magnon dispersion along the out-of-plane direction has a bandwidth of less than 20 meV indicating the dominant spin-spin interactions are within the \kag{}-lattice planes. The nearly two-dimensional character of the spin excitation spectrum is further evidenced by the rod-like scattering shown in Fig.~\ref{fig:lowE_spec}(c).

\begin{table*}[t]
\caption{\label{tab:paras}%
Hamiltonian parameters determined from the spin wave theory analysis. 
}
\begin{ruledtabular}
{\renewcommand{\arraystretch}{1.2}
\begin{tabular}{ccccccccc}
Label (number of paths) & $J_1$ (4)   & $J_{int1}$ (2)  & $J_{2}$ (4)    & $J_{int2}$ (8)  & $J_{3}$ (2) & $J_{4}$ (4)  & $D_{z}$ \\ \hline
$J_{ij}^\text{Fit}$ (meV)       &  -44.33 $\pm1.56$   &    4.51$ \pm1.00$     &    12.23 $\pm1.06$     &   1.27 $\pm0.24$     &  -5.28 $\pm2.32$   &   -4.60 $\pm0.90$   &    0.1    \\
Distance (\AA) &  2.65   &    4.45     &  4.59      &  5.18       &   5.30  & 5.30    & -    \\
\end{tabular}
}  
\end{ruledtabular}
\end{table*}


The high-energy spectra were measured using the SEQUOIA  spectrometer with $E_i$=500 meV. We integrate the INS data over $-4\leq L \leq$4~r.l.u. to enhance statistics. Note that due to momentum and energy conservation, high-energy transfer data is obtained from a larger magnitude $L$-region, which results in lower scattering intensity from the magnetic form factor contribution. 
As shown in Fig.~\ref{fig:highE_spec}, the excitations extend to at least 200~meV. Two individual magnon branches are observed corresponding to the lower- and mid-magnon bands in Fig.\ref{fig:crystallography}(c) of the ferromagnetic \kag{} spin-lattices through the M- and K-points in the BZ. The higher energy spectral weight above $\sim$120 meV is diffuse, and becomes  indiscernible from background above $\sim$200 meV. Figure~\ref{fig:highE_spec}(c) shows momentum scans through $\Gamma$'-$M$-$\Gamma$-K-X for increasing energy transfer.  Along both the $\Gamma$-M and $\Gamma$-K directions, the peak linewidths  broaden as a function of $\textbf{Q}$ near the zone boundary (ZB), and the peak-positions are intact over a wide energy range 120$<E<$170 meV (80$<E<$120 meV) near the K (M)-point. These $\textbf{Q}$-, $E$- peak broadenings indicate the decay of the magnons, resulting from the quasiparticle scattering~\cite{zhitomirsky2013,oh2016,chen2020hund}. Considering the metallicity of FeSn along with the collinear spin configuration, FeSn presumably has a large magnon-electron interaction, which results in strong damping of the magnon spectra. 


\begin{figure*}[t]
\includegraphics[width=16.5cm]{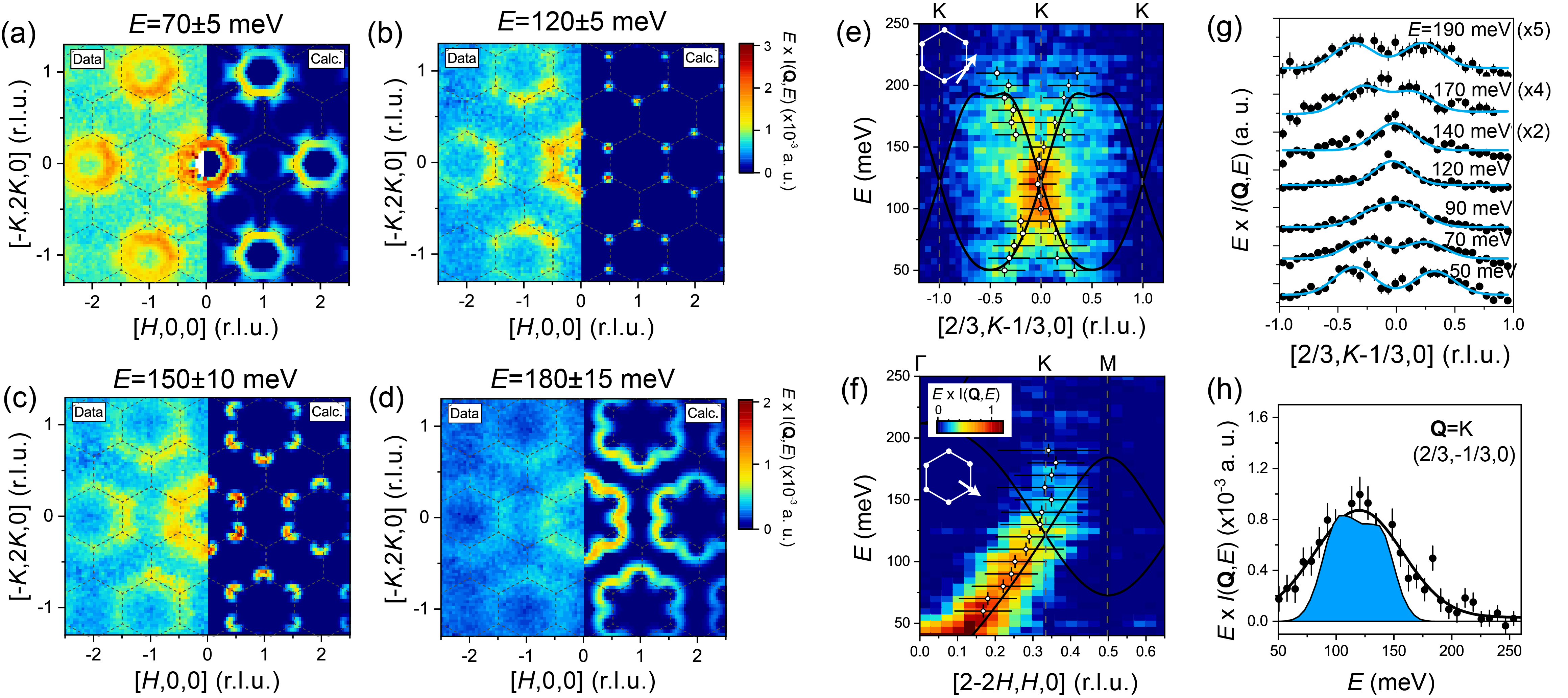} 
\caption{\label{fig:dirac} 
(a)-(d) Constant energy slices of the INS data ($E\times I(\text{\textbf{Q}},E)$) and spin wave calculations. Dashed lines indicate the first BZ in the $HK$-plane. The color bar for (a)(b) ((c),(d)) is shown in right of (b) ((d)). INS spectra through the K-point along (e) transverse- and (f) radial-directions (see arrows in insets). (g) Momentum scans at constant energy through the K-point along the transverse direction. The dispersion was extracted by fitting the spectra to Gaussian functions (solid lines) and the results are displayed as circles in (e). 
The lines in (e)(f) represent the linearly crossing magnons for $L$=0.5 at the Dirac node. Horizontal (vertical) error bars in (e)(f) indicate the fitted FWHM (range of energy integration). (h) Constant wave-vector scan at the Dirac point.  Data are shown as symbols and the spectral weight from LSWT (shaded region) is described in the text. The line is a guide to eye. The data was obtained by integrating over the momentum region [$H$,0,0]=$\pm$0.05, [2$K$,-$K$,0]=$\pm$0.06, and [0,0,$L$]=$\pm$4. (e),(f),(h) For clarity, the nonmagnetic background was obtained from the scattering at $\textbf{Q}$=(2/3,2/3,0) and subtracted from the measured intensities~\cite{supple}.
}
\end{figure*}

To understand the observed spin wave spectra and the underlying spin-spin interactions, we use linear spin wave theory (LSWT) with the Hamiltonian, ${\cal H}=J_{n}\sum_{i,j} S_{i}S_{j} -D_{z}\sum_{i}(S_{i}^{z})^{2}$, as implemented in the SpinW software package~\cite{toth2015}. We set $S$=1 considering the measured effective moment of 3.4 $\mu_{B}$ (2.8 $\mu_{B}$ for $S$=1, where $g$=2)~\cite{sales_2019}. $J_{n}$ and $D_{z}$ correspond to Heisenberg exchange couplings for the $n$th nearest-neighbor and a single-ion anisotropy, respectively~\footnote{Indeed, Dzyalloshinskii-Moriya interaction (DMI) along $z$-axis ($c$) is symmetrically allowed. However, since the DMI on the spins aligning in the plane ($z\perp S$) does not change the dispersion we exclude the DMI in the model Hamiltonian.}. Interactions up to fourth (second) nearest-neighbor in-the-plane (out-of-plane) direction (see Fig.~\ref{fig:crystallography}(a)(b)) were considered. Note that $J_3$ and $J_4$ have the same distance but different paths. Hence, the distinction of these parameters is maintained due to the potential effects on the RKKY interaction of the complicated band structure near the fermi surface~\cite{roth1966}. 
The measured dispersion is fitted to the calculated dispersion (see \cite{supple}), yielding the parameters listed in  Table~\ref{tab:paras}. The  parameters indicate a dominant nearest-neighbor ferromagnetic interaction $J_{1}$ responsible for the ferromagnetic \kag{} spin-lattice.  We also determine non-negligible further neighbor exchanges, $J_2 (\sim-0.28J_1)$, $J_3 (\sim0.12J_1)$, and $J_4 (\sim0.1J_1)$, are present. The sign and relative size of the parameters from the spin wave analysis are largely consistent with parameters determined from first principles calculations (see Supplemental materials~\cite{supple} and \cite{sales_2019}). Furthermore, the symmetry-allowed easy-plane single-ion anisotropy ($D_{z}>$0) reproduces the peaked intensity data near 4 meV shown in Fig~\ref{fig:lowE_spec}(d)~\cite{supple}. 

The refined spin wave scattering intensity is compared to the experimental data in Figs.~\ref{fig:lowE_spec},~\ref{fig:highE_spec}, and ~\ref{fig:dirac}. The calculations reproduce the low-energy spectra. However, the scattering intensity and dispersion  deviate from the calculation at the zone boundary and well-defined modes are essentially absent above 200 meV in the measurements. This discrepancy in the scattering intensity is ascribed to interactions with the Stoner continuum~\cite{korenman1972,ibuka2017,Adams2000,diallo2009}, and indicates the energy scale of the Stoner excitations. Due to the large number of electronic bands in FeSn, it is challenging to make direct comparisons to the magnetic spectra measured here.  However, electronic band structure calculations do indicate splitting of majority and minority spin bands near the fermi energy~\cite{lin2020,sales_2019}. The minimum energy of an indirect inter-band transition for these bands near $\textbf{Q}$=$\Gamma$ is $\sim$0.1-0.2 eV, which results in a gap of the Stoner excitations with finite momenta (see Fig.\ref{fig:crystallography}(e)), and is consistent with the energy scale above which damping begins to dominate the INS spectra. 


The determined spin Hamiltonian and the symmetry of the spin configuration preserves time reversal symmetry, and permit the existence of a Dirac point in the magnon spectrum. LSWT presents a sharp linear magnon band crossing at $E\sim$120 meV at the K-point (see dispersion line in Fig.~\ref{fig:highE_spec}(a)(b)). However, due to interactions with electron, the spin waves near the Dirac node are susceptible to decay. Figure~\ref{fig:dirac}(a)-(d) presents constant energy slices measured up to 180 meV. The low-energy spectrum below the Dirac node are reproduced by LSWT.  The Dirac node is evident at the K-point at 120 meV as shown in Fig.~\ref{fig:dirac}(b).  Above  120 meV, the excitations  significantly broaden.  The is particularly evident near the zone boundary and the broadening increases with increasing energy transfer.   Figures~\ref{fig:dirac}(e) and (f) highlight the dispersion in the vicinity of the Dirac nodes along transverse and radial directions, respectively.  Figure~\ref{fig:dirac}(g) shows constant energy scans along the transverse direction through the Dirac nodal point as having two  clear peaks below 100 meV and above 150 meV, but only a single peak between 100 meV and 150 meV in the vicinity of the two crossing bands.  Peak positions  extracted from Gaussian fits compare well to the LSWT dispersion curve in Figs.~\ref{fig:dirac}(e).
We note that finite spectral weight likely due to damping from interactions with the continuum is present between the two peaks above the Dirac node.
In contrast, the momentum scan along the radial direction deviates from the calculated dispersion above 120 meV, as shown in Fig.~\ref{fig:dirac}(f). Rather than two peaks, constant energy scans along this direction show a broadened spectral weight centered near the Dirac node. These results demonstrate that the scattering with itinerant electrons reconstruct the upper Dirac cone dispersion, but also the diffusive continuum from the decay fills in the Dirac cone. 
Figure~\ref{fig:dirac}(h) shows an energy scan at the Dirac node compared to the calculated spectral weight of the LSWT model convoluted with the instrumental energy resolution. The decayed spectral weight is visible above 150 meV and extends well beyond the LSWT model of the scattering. 



Additionally, the LSWT completely fails to explain the observed upper spectral weight above 120 meV along $\Gamma$ to M (see Fig~\ref{fig:highE_spec}(a)(b)). We note that adjusting the exchange values of the Hamiltonian to have a large antiferromagnetic $J_4 \sim$3.5 meV (with ferromagnetic $J_3$) decreases the upper magnon branch down to 120 meV and reproduces the observed spectral weight near M. 
However, this results in a large discrepancy in the other magnon bands (see Supplemental Material~\cite{supple}). 
It is worth noting that a fluctuation continua is also present at the top of the lower magnon branch at $\textbf{Q}$=M (zone boundary) above 80 meV (see Fig.~\ref{fig:highE_spec}(d)). It connects the lower magnon branch to the upper spectral weight without a gap in the spectrum.  This in turn generates a band touching at M around the Dirac node, resulting in a weak ring-shaped spectral weight in the all constant energy slices between 80 meV and 150~meV (see Fig.~\ref{fig:dirac}). This continuous scattering confirms that the excitation near M is not simply due to a spin wave excitation. 
Therefore, a likely component of the measured spectral weight near M is the decayed spectra of the upper magnon band. To explain this may require a comparison to the itinerant band model~\cite{ewings2011,diallo2009}, a more sophisticated approach which includes the correction from the interactions with itinerant electrons~\cite{park2011,muller2016}, or spin-fermion model~\cite{spinfermion1,spinfermion2,spinfermion3}.

In summary, we have found that the spin excitation spectrum in the ferromagnetic \kag{} metal FeSn is quasi-two-dimensional with progressively stronger damping of the spin waves with increasing energy transfer. The determined exchange terms for the spin Hamiltonian  provide for a  symmetry allowed magnon Dirac nodal point near the electronic continua. The interaction with the itinerant electrons is large near the nodal point, resulting in a significant spectral broadening with momentum dependence. The interactions are also large near the M-point, which results in continuous spectral weight between the lower and upper magnon bands. A more complete understanding of these observations require calculations which account for the electron-magnon interactions. It will be particularly interesting to check if the spin-charge coupled spectra in the \kag{} metallic magnet possesses the topology arising from correlation effects.

\begin{acknowledgments}
We acknowledge M. Lumsden for useful discussions. This research was supported by the U.S. Department of Energy, Office of Science, Basic Energy Sciences, Materials Science and Engineering Division. Work at the Oak Ridge National Laboratory Spallation Neutron Source was supported by U.S. DOE, Office of Science, BES, Scientific User Facilities Division. The neutron experiment at the Materials and Life Science Experimental Facility of the J-PARC was performed under a user program (Proposal No. 2019B0248 and 2020A0217).
\end{acknowledgments}


\providecommand{\noopsort}[1]{}\providecommand{\singleletter}[1]{#1}%

\end{document}